\documentclass[graybox]{svmult}

\usepackage{mathptmx}       %
\usepackage{helvet}         %
\usepackage{courier}        %
\usepackage{type1cm}        %
\usepackage{makeidx}         %
\usepackage{graphicx}        %
\usepackage{multicol}        %
\usepackage[bottom]{footmisc}%

\usepackage{amsmath}
\usepackage[OT1]{fontenc}
\PassOptionsToPackage{hyphens}{url}  %
\usepackage[hidelinks]{hyperref}  %

\makeindex             %

\begin{document}

\title*{``Guess Who ?'' Large-Scale Data-Centric Study of the Adequacy of Browser Fingerprints for Web Authentication}
\titlerunning{Data-Centric Study of the Adequacy of Browser Fingerprints for Web Authentication}
\author{Nampoina~Andriamilanto, Tristan~Allard, and Ga\"etan~Le~Guelvouit}
\institute{Nampoina~Andriamilanto \at Institute of Research and Technology b$<>$com, Cesson-Sévigné, France \email{nampoina.andriamilanto@b-com.com}
\and Tristan~Allard \at Univ Rennes, CNRS, IRISA, Rennes, France \email{tristan.allard@irisa.fr}
\and Ga\"etan~Le~Guelvouit \at Institute of Research and Technology b$<>$com, Cesson-Sévigné, France \email{gaetan.leguelvouit@b-com.com}}
\maketitle

\abstract{
  Browser fingerprinting consists in collecting attributes from a web browser to build a browser fingerprint.
  In this work, we assess the adequacy of browser fingerprints as an authentication factor, on a dataset of $4,145,408$~fingerprints composed of $216$~attributes.
  It was collected throughout $6$~months from a population of general browsers.
  We identify, formalize, and assess the properties for browser fingerprints to be usable and practical as an authentication factor.
  We notably evaluate their distinctiveness, their stability through time, their collection time, and their size in memory.
  We show that considering a large surface of $216$~fingerprinting attributes leads to an unicity rate of $81$\% on a population of $1,989,365$~browsers.
  Moreover, browser fingerprints are known to evolve, but we observe that between consecutive fingerprints, more than $90$\% of the attributes remain unchanged after nearly $6$~months.
  Fingerprints are also affordable.
  On average, they weigh a dozen of kilobytes, and are collected in a few seconds.
  We conclude that browser fingerprints are a promising additional web authentication factor.
}

\abstract*{
  Browser fingerprinting consists in collecting attributes from a web browser to build a browser fingerprint.
  In this work, we assess the adequacy of browser fingerprints as an authentication factor, on a dataset of $4,145,408$~fingerprints composed of $216$~attributes.
  It was collected throughout $6$~months from a population of general browsers.
  We identify, formalize, and assess the properties for browser fingerprints to be usable and practical as an authentication factor.
  We notably evaluate their distinctiveness, their stability through time, their collection time, and their size in memory.
  We show that considering a large surface of $216$~fingerprinting attributes leads to an unicity rate of $81$\% on a population of $1,989,365$~browsers.
  Moreover, browser fingerprints are known to evolve, but we observe that between consecutive fingerprints, more than $90$\% of the attributes remain unchanged after nearly $6$~months.
  Fingerprints are also affordable.
  On average, they weigh a dozen of kilobytes, and are collected in a few seconds.
  We conclude that browser fingerprints are a promising additional web authentication factor.
}

\section{Introduction}
  Web authentication widely relies on identifier-password pairs.
  Passwords are easy to use, but suffer from severe security flaws.
  Indeed, users use common passwords, paving the way to brute-force or guessing attacks~\cite{BON12}.
  They also reuse passwords across websites~\cite{WJHBW18} which increases the impact of attacks.
  Phishing attacks are also a major threat to passwords.
  Over the course of a year, Thomas et al.~\cite{TLZBRIMCEMMPB17} achieved to retrieve $12.4$~million credentials stolen by phishing kits.
  These flaws gave rise to multi-factor authentication~\cite{BHOS15}, such that each additional authentication factor provides an \emph{additional security barrier}.
  However, this usually comes at the cost of \emph{usability} (i.e., users have to remember, possess, or do something).

\begin{figure}
    \centering
    \includegraphics[width=.7\columnwidth]{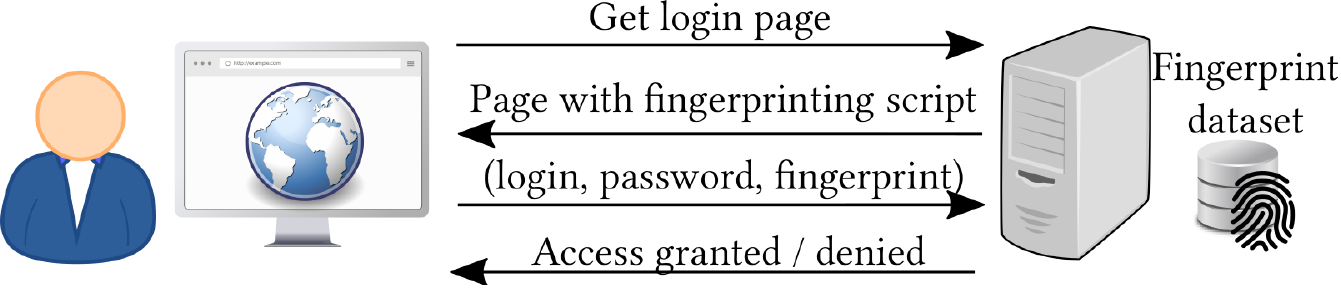}
    \caption{Simplified web authentication mechanism relying on browser fingerprinting.}
    \label{fig:auth-mechanism-scheme}
  \end{figure}

  In the meantime, \emph{browser fingerprinting} gains attention.
  The seminal Panopticlick study~\cite{ECK10} highlights the possibility to build a \emph{browser fingerprint} by collecting attributes from a web browser.
  In addition to being widely used for web tracking purposes~\cite{EN16} (raising legal and ethical issues), browser fingerprints are used as an authentication factor \emph{in real-life}.
  Browser fingerprints are indeed a \emph{good candidate} as an authentication factor thanks to their distinctive power, their frictionless deployment (e.g., no additional software), and their usability (no secret to remember, no additional object to possess, and no supplementary action to carry out).
  As a result, companies like MicroFocus\footnote{
    \url{https://www.netiq.com/documentation/access-manager-44/admin/data/how-df-works.html}
  } or SecureAuth\footnote{
    \url{https://docs.secureauth.com/pages/viewpage.action?pageId=33063454}
  } include browser fingerprints within their authentication mechanisms (see Figure~\ref{fig:auth-mechanism-scheme} for an example of such mechanism).

  \textbf{Related works.}
  To the best of our knowledge, no large-scale study rigorously evaluates the adequacy of browser fingerprints as an authentication factor.
  Most works about their use for authentication concentrate on the design of an authentication mechanism~\cite{UMFHSW13, PJ15, LABN19, REKP19}, and the empirical studies on browser fingerprints focus on their efficacy as a web tracking tool~\cite{ECK10, LRB16, GLB18}.
  Such a mismatch between the understanding of browser fingerprints for authentication -- currently poor -- and their ongoing adoption in real-life is a serious harm to the security of web users.
  The lack of documentation from the existing tools (e.g., about the used attributes, about the distinctiveness of the resulting fingerprints, about their stability) only adds up to the current state of ignorance.
  All this whereas security-by-obscurity contradicts the most fundamental security principles.

  \textbf{Our contributions.}
  We conduct the first \emph{large-scale data-centric empirical study of the fundamental properties} of browser fingerprints when used as an additional authentication factor.
  We base our findings on an in-depth analysis of a real-life fingerprint dataset collected over $6$~months, that contains $4,145,408$~fingerprints composed of $216$~attributes.
  We formalize, and assess on our dataset, the properties necessary for paving the way to elaborate browser fingerprinting authentication mechanisms.
  These properties are usually used to evaluate biometric characteristics~\cite{MMJP03}.
  We stress that we do not make any assumption on the inner working of the authentication mechanism, and consequently on the adversarial strategy.
  Our properties aim at characterizing the adequacy and the practicability of browser fingerprints, independently of their use within future authentication mechanisms.
  In particular, we measure the size of browser anonymity sets through time, and show that $81$\% of our fingerprints are unique.
  Moreover, we measure the proportion of identical attributes between two observations of the fingerprint of a browser, and show that $90$\% of the attributes remain unchanged after nearly $6$~months.
  Finally, we measure the collection time and the size of fingerprints.
  We show that on average, they weigh a dozen of kilobytes, and are collected in a few seconds.

  The rest of the paper is organized as follows.
  Section~\ref{sec:authentication-factor-properties} presents and formalizes the properties evaluated in our analysis.
  Section~\ref{sec:dataset} describes the analyzed dataset.
  Section~\ref{sec:results} presents the experimental results.
  Finally, Section~\ref{sec:conclusion} synthesizes the results and concludes.

\section{Authentication Factor Properties}
\label{sec:authentication-factor-properties}
  The ``Handbook of Fingerprint Recognition''~\cite{MMJP03} summarizes the properties that a biometric characteristic requires to be \emph{usable}\footnote{
    Here, \emph{usable} refers to the adequacy of the characteristic to be used for authentication, rather than the ease of use by the users.
  } as an authentication factor, and the additional properties required for a biometric authentication scheme to be \emph{practical}.
  We make the link between the fingerprints used to recognize persons, and the ones used to recognize browsers.
  Therefore, we evaluate browser fingerprints according to these properties to assert their adequacy for web authentication.
  In this section, we list these properties, formalize how to measure some, and explain why the others are not addressed in this study.

  The four properties needed for a biometric characteristic to be usable as an authentication factor are the following.
  \begin{itemize}
    \item \emph{Universality}: the characteristic should be present in everyone.
    \item \emph{Distinctiveness}: two distinct persons should have different characteristics.
    \item \emph{Permanence}: the same person should have the same characteristic over time. We rather use the term \emph{stability}.
    \item \emph{Collectibility}: the characteristic should be collectible and measurable.
  \end{itemize}

  The three properties that a biometric authentication scheme requires to be practical are the following.
  \begin{itemize}
    \item \emph{Performance}: the scheme should consume few resources, and be robust against environmental changes.
    \item \emph{Acceptability}: the users should accept to use the scheme in their daily lives.
    \item \emph{Circumvention}: it should be difficult for an attacker to deceive the scheme.
  \end{itemize}

  The properties that we study are the \emph{distinctiveness}, the \emph{stability}, and the \emph{performance}.
  We consider that the \emph{universality} and the \emph{collectibility} are satisfied, as the HTTP headers that are automatically sent by browsers constitute a fingerprint.
  However, we stress that a loss of distinctiveness occurs when no JavaScript attribute is available.
  About the \emph{circumvention}, we refer the reader to Laperdrix et al.~\cite{LABN19} that analyzed the security of an authentication mechanism based on browser fingerprints.
  We let the evaluation of the \emph{acceptability} as future works, but we stress that such mechanisms are already used in a rudimentary form\footnote{
    \url{https://support.google.com/accounts/answer/1144110}
  }.

  \subsection{Distinctiveness}
    To satisfy the \emph{distinctiveness} property, the browser fingerprints should distinguish two different browsers.
    The distinctiveness depends on the used attributes, and on the fingerprinted browser population.
    The two extreme cases are every browser sharing the same fingerprint, which makes them indistinguishable from each other, and no two browsers sharing the same fingerprint, making every browser distinguishable.

    Our dataset entries are composed of a fingerprint, the source browser, and the time of collection in the form of a Unix timestamp in milliseconds.
    We denote~$B$ the domain of the unique identifiers, and~$T$ the timestamp domain.
    The fingerprint dataset is denoted~$D$, and is formalized as:
    \begin{equation}
      D = \{ (f, b, t) \mid f \in F, b \in B, t \in T \}
    \end{equation}

    We use the size of the browser anonymity sets to quantify the distinctiveness, as the browsers that belong to the same anonymity set are indistinguishable.
    We denote~$S(f, D)$ the function that returns the browsers that provided the fingerprint~$f$ in the dataset~$D$.
    It is formalized as:
    \begin{equation}
      S(f, D) = \{
        b \in B \mid
        \forall (g, b, t) \in D, f = g
      \}
    \end{equation}

    We denote~$A(\epsilon, D)$ the function that provides the fingerprints that have an anonymity set of size~$\epsilon$ (i.e., that are shared by $\epsilon$~browsers) in the dataset~$D$.
    It is formalized as:
    \begin{equation}
      A(\epsilon, D) = \{
        f \in F \mid
        \mathrm{card}(S(f, D)) = \epsilon
      \}
    \end{equation}

    We measure the anonymity set sizes on the fingerprints currently in use by each browser, and not on their whole history.
    It is performed by simulating datasets composed of the last fingerprint seen for each browser at a given time.
    Let~$E_{\tau}(D)$ be the simulated dataset originating from~$D$ that represents the state of the fingerprints after~${\tau}$~days.
    With~$t_{\tau}$~the last timestamp of this day, we have:
    \begin{equation}
      E_{\tau}(D) = \{
        (f_i, b_j, t_k) \in D \mid
        \forall (f_p, b_q, t_r) \in D,
        b_j = b_q,
        t_r \leq t_k \leq t_{\tau}
      \}
    \end{equation}

  \subsection{Stability}
    Browser fingerprints have the particularity of evolving through time, due to changes in the web environment like a software update or a user configuration.
    We measure the \emph{stability} by the mean similarity between two consecutive fingerprints observed for a browser, given the elapsed time between them.
    The two extreme cases are every browser holding the same fingerprint through its life, and the fingerprint of a browser changing completely between each observation.

    We denote~$C(\Delta, D)$ the function that provides the consecutive fingerprints of $D$ that are separated by a time lapse comprised in the~$\Delta$ time range.
    It is formalized as:
    \begin{equation}
      \begin{aligned}
        C(\Delta, D) = \{ (f_i, f_p) \mid \, &
          \forall ((f_i, b_j, t_k), (f_p, b_q, t_r)) \in D^2, \\
          & b_j = b_q, t_k < t_r, (t_r - t_k) \in \Delta \}
      \end{aligned}
    \end{equation}

    We consider the Kronecker delta~$\delta(x, y)$, being~$1$ if~$x$ equals~$y$, and~$0$ otherwise.
    We consider the set $\Omega$ of the $n$ used attributes.
    We denote~$f[\omega]$ the value taken by the attribute~$\omega$ for the fingerprint~$f$.
    Let~$\mathrm{sim}(f, g)$ be a simple similarity function between the  fingerprints $f$ and $g$, which is formalized as:
    \begin{equation}
      \mathrm{sim}(f, g) = \frac{1}{n}
      \sum_{\omega \in \Omega} \delta(f[\omega], g[\omega])
    \end{equation}

    We define the function~$\mathrm{meansim}(\Delta, D)$ that provides the mean similarity of the consecutive fingerprints, for a given time range~$\Delta$ and a dataset~$D$.
    It is formalized as:
    \begin{equation}
      \mathrm{meansim}(\Delta, D) =
        \frac{
          \sum_{(f, g) \in C(\Delta, D)}
            \mathrm{sim}(f, g)
        }{
          \mathrm{card}(C(\Delta, D))
        }
    \end{equation}

  \subsection{Performance}
    We consider three aspects of the performance of browser fingerprints for authentication.
    They are the collection time of the fingerprints, their size in memory, and the loss of efficacy between the fingerprints of different device types.

    We measure the \emph{collection time} of our fingerprints composed of $200$~JavaScript attributes, and ignore the HTTP headers that are transmitted passively.
    We measure the \emph{size} of our fingerprints, having the canvas images~\cite{MS12} stored as sha256 hashes.
    We stress that compressing the fingerprint to a single hash is unpractical due to the evolution of fingerprints.
    Previous works showed that \emph{mobile} and \emph{desktop devices} present differences in the properties of their browser fingerprints~\cite{SPJ15, LRB16, GLB18}, notably that mobile browsers have less distinctive fingerprints.
    Following these findings, we assess that the properties of the fingerprints of mobile and desktop browsers are similar.

\section{Fingerprint Dataset}
\label{sec:dataset}
  To study the properties of browser fingerprints on a real-world browser population, we launched a fingerprint collection experiment.
  It was performed in collaboration with the authors of~\cite{GLB18}, and an industrial partner that controls one of the top $15$~French websites according to Alexa\footnote{
    \url{https://www.alexa.com/topsites/countries/FR}
  }.
  The authors of~\cite{GLB18} held the $17$ attributes of their previous work~\cite{LRB16} and focused on web tracking, whereas we held $216$ attributes and focused on web authentication.

  \begin{table*}
    \centering
    \caption{
      Dataset comparison between Panopticlick, AmIUnique, Hiding in the Crowd, and this study.
      \textbf{-} denotes missing information, and \textbf{*} denotes deduced information.
      The attributes only comprise the original ones, and the fingerprints are counted after preprocessing.
    }
    \label{tab:unicity_comparison}
    \begin{tabular}{|c|c|c|c|c|}
      \hline
                            & PTC~\cite{ECK10} & AIU~\cite{LRB16}   & HITC~\cite{GLB18} & \textbf{This study} \\
      \hline
        Collection period   & 3 weeks      & 3-4 months* & 6 months              & \textbf{6 months}        \\
        Number of attributes             & 8            & 17          & 17               & \textbf{216}       \\
        Number of browsers            & -            & -           & -                & \textbf{1,989,365}        \\
        Number of fingerprints        & 470,161      & 118,934     & 2,067,942           & \textbf{4,145,408}        \\
        Number of distinct fingerprints        & 409,296      & 142,023\footnotemark     & -           & \textbf{3,578,196}        \\
        Proportion of desktop fingerprints   & -            & 0.890*      & 0.879               & \textbf{0.805}       \\
        Proportion of mobile fingerprints    & -            & 0.110*      & 0.121               & \textbf{0.134}       \\
        Unicity of global fingerprints      & 0.836        & 0.894       & 0.336               & \textbf{0.818}       \\
        Unicity of mobile fingerprints     & -            & 0.810       & 0.185               & \textbf{0.399}       \\
        Unicity of desktop fingerprints    & -            & 0.900       & 0.357               & \textbf{0.884}       \\
      \hline
    \end{tabular}
  \end{table*}

  \footnotetext{
    This number is provided in Figure~$11$ as the distinct fingerprints, but also corresponds to the raw fingerprints.
    Every fingerprint would be unique if the number of distinct and collected fingerprints are equal, hence we are not confident in this number, but it is the one provided by the authors.
  }

  \subsection{Fingerprint Collection}
    We designed a fingerprinting probe that collects $200$~JavaScript properties and $16$~HTTP header fields.
    We integrated the probe to two general audience web pages of our industrial partner, which subjects are political news and weather forecast.
    The probe collected fingerprints from December $7$, $2016$, to June $7$, $2017$.
    Only the visitors that consented to cookies were fingerprinted, in compliance with the European directives 2002/58/CE and 2009/136/CE in effect at the time.
    To differentiate browsers, we assigned them a unique identifier (UID) as a $6$-months cookie.
    Similarly to~\cite{ECK10, LRB16}, we coped with cookie deletion by storing a one-way hash of the IP address, computed by a secure cryptographic hash function.

    Previous datasets were collected through dedicated websites, and are biased towards privacy-aware and technically-skilled persons~\cite{ECK10, LRB16}.
    Our population is more general audience oriented, but the website audience is mainly French-speaking users.
    This leads to a bias towards this population.
    The timezone is set to $-1$ for $98.48$\% of browsers, $98.59$\% of them have daylight saving time enabled, and \texttt{fr} is present in $98.15$\%~of the \texttt{Accept-Language} HTTP header.

  \subsection{Dataset Filtering and Preprocessing}
    Given the experimental aspect of fingerprints and the scale of our collection, the raw dataset contained erroneous or irrelevant samples.
    We remove $70,460$~entries entries that have a wrong format (e.g., empty or truncated data), that are duplicated, or that come from a robot.

    Cookies are an unreliable identification method, hence we perform a resynchronization similar to~\cite{ECK10}.
    We consider the entries that have the same (fingerprint, IP address hash) pair to come from the same browser, and assign them the same UID.
    Similarly to~\cite{ECK10}, we do not synchronize the interleaved UIDs, that are the pairs having the UID values $b_1$, $b_2$, then $b_1$ again.
    We replace $181,676$~UIDs with $116,708$ replacement UIDs using this method.

    To avoid counting multiple entries of identical fingerprints coming from the same browser, the usual way is to ignore them during collection~\cite{ECK10, LRB16}.
    Our probe collects the fingerprint on each visit, and to stay consistent with common methodologies we deduplicate the fingerprints afterward.
    For each browser, we hold the first entry that has a given fingerprint, and ignore the consecutive entries if they have this fingerprint.
    For example, if a browser $b$ has the entries $\{(f_1, b, t_1), (f_2, b, t_2), (f_2, b, t_3), (f_1, b, t_4)\}$, we hold $\{(f_1, b, t_1), (f_2, b, t_2), (f_1, b, t_4)\}$.
    The deduplication constitutes the biggest cut in our dataset, with $2,420,217$ entries filtered out.

    We extract $46$~additional attributes from $9$~original attributes, which are of two types.
    The first type consists in extracted attributes composed of parts of their original attribute, like the screen resolution that is split into the values of the width and the height.
    The second type consists of information sourced from an original attribute,
    like the number of plugins extracted from the list of plugins.

  \subsection{Work Dataset}
  \label{sec:work-dataset}
    The work dataset obtained after the preprocessing step contains $5,714,738$~entries (comprising identical fingerprints for a given browser if they are interleaved), with $4,145,408$~fingerprints (no identical fingerprint counted for the same browser), and composed of $3,578,196$~distinct fingerprints.
    The fingerprints are composed of $216$~original attributes and $46$~extracted ones, for a total of $262$~attributes.
    They come from $1,989,365$~browsers, $27.53$\%~of which have multiple fingerprints.
    Table~\ref{tab:unicity_comparison} presents a comparison between the dataset of Panopticlick~\cite{ECK10}, AmIUnique~\cite{LRB16}, Hiding in the Crowd~\cite{GLB18}, and this study.

\section{Empirical Evaluation of Browser Fingerprints Properties}
\label{sec:results}
  In this section, we evaluate the browser fingerprints of our dataset according to the properties presented in Section~\ref{sec:authentication-factor-properties}.
  We show that our fingerprints offer a satisfying \emph{distinctiveness}, as $81$\% of them are only shared by one browser.
  Moreover, our fingerprints are \emph{stable}, as more than $90$\%~of the attributes are expected to stay identical between two observations, even if they are separated by nearly $6$~months.
  Our fingerprints of \emph{mobile browsers} are less distinctive than the fingerprints of desktop browsers, with a unicity rate of $42$\% against $84$\%.
  Finally, our fingerprints do not hinder the \emph{performance} as, on average, they weigh a dozen of kilobytes and take a few seconds to collect.

  \subsection{Distinctiveness}
    We call unicity rate the proportion of the fingerprints that are shared by a single browser.
    Our fingerprints offer a satisfying distinctiveness, as they have a stable unicity rate of approximately $81$\% on the long run, and more than $94$\% of the fingerprints are shared by $8$ browsers or less.
    However, the fingerprints of the mobile browsers are more uniform than the fingerprints of the desktop browsers, with a unicity rate of approximately~$42$\% against $84$\%, on the long run.

    \begin{figure}
      \centering
      \includegraphics[width=.7\columnwidth]{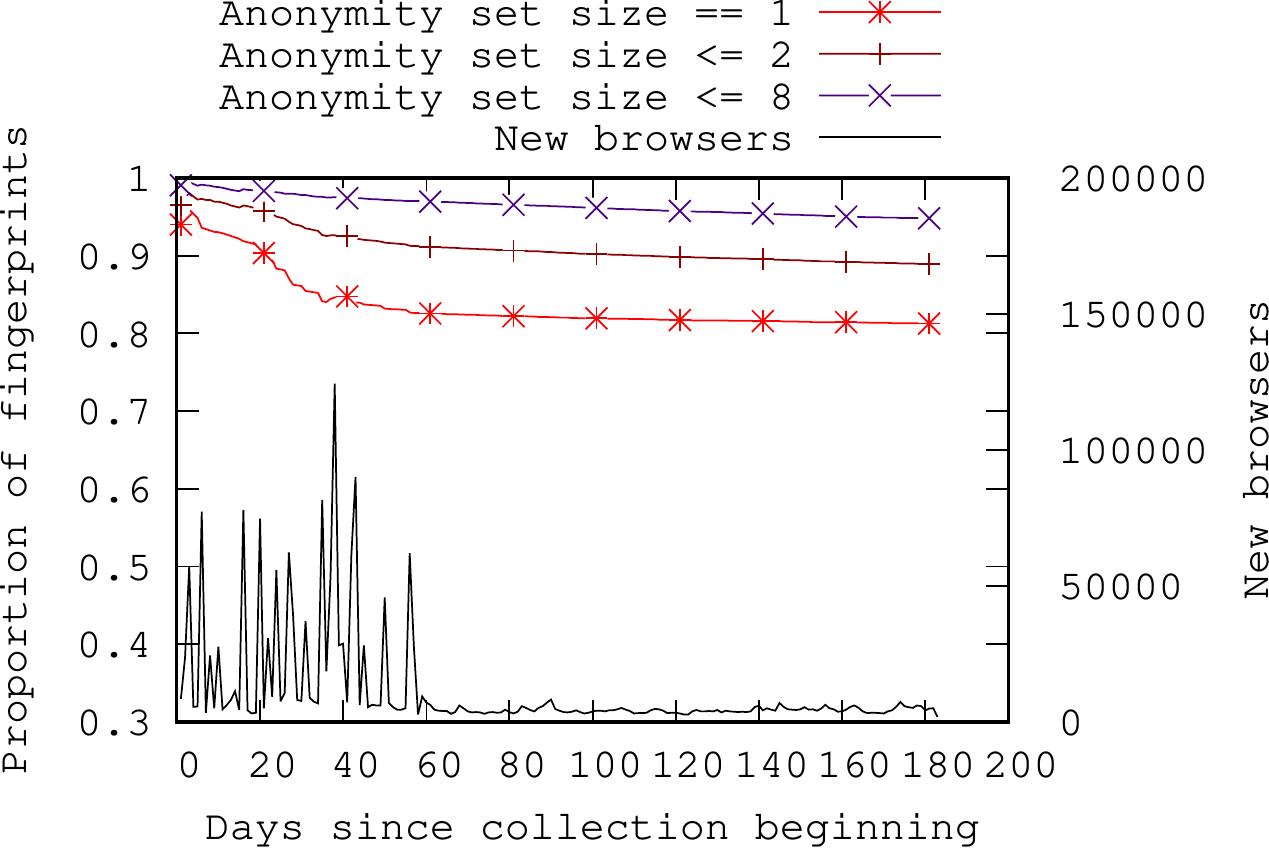}
      \caption{Anonymity set sizes and frequency of browser arrivals through the partitioned datasets obtained after each day.}
      \label{fig:anonymity-sets-full}
    \end{figure}

    Figure~\ref{fig:anonymity-sets-full} presents the size of the anonymity sets (AS) alongside the frequency of browser arrival for the daily-partitioned datasets.
    New browsers are encountered continually, but starting from the $60$th~day, the arrival frequency stabilizes around $5,000$~new browsers per day.
    Before this stabilization, we have a variable arrival frequency with some major spikes that seem to correspond to events that happened in France.
    For example, the spike on the $38$th day corresponds to a live political debate on TV, and the spike on the $43$rd day correlates with the announcement of a cold snap.

    \begin{figure}
      \centering
      \includegraphics[width=.7\columnwidth]{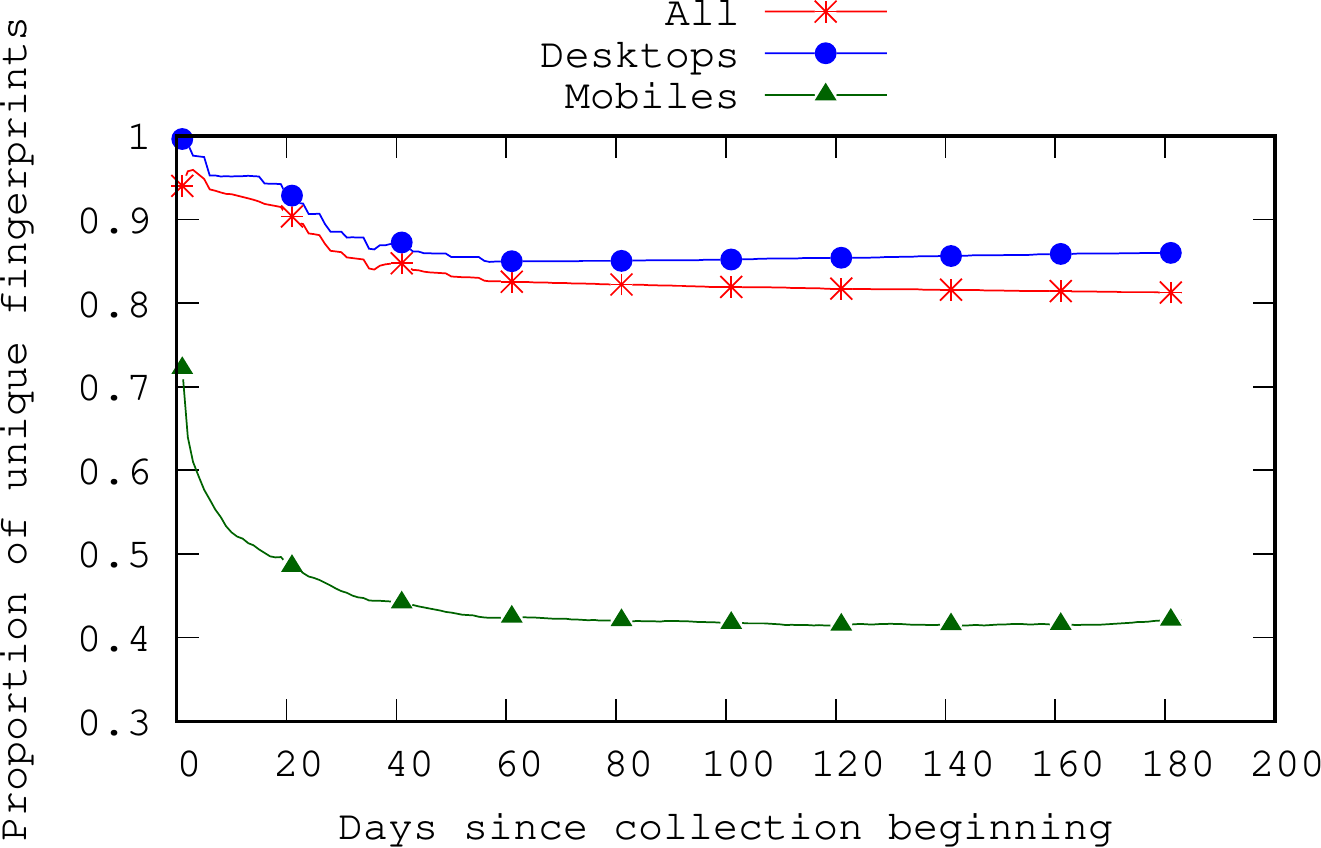}
      \caption{Proportion of unique fingerprints for the overall, the mobile, and the desktop groups, through the partitioned datasets obtained after each day.}
      \label{fig:anonymity-sets-web-environment}
    \end{figure}

    Figure~\ref{fig:anonymity-sets-web-environment} presents the unicity rate through the partitioned datasets for the overall, the mobile, and the desktop groups.
    The unicity rate is stable for the desktop browsers, with a slight increase of~$1.04$~points from the~$60$th day to the~$183$th, from~$84.99$\% to~$86.03$\%.
    The unicity rate of the mobile browsers is lower, and it has a little decrease of~$0.29$~points on the same period, from~$42.42$\% to~$42.13$\%.

  \subsection{Stability}
    Our fingerprints are stable, as a browser is expected to have at least~$90$\%~of its attributes unchanged, even after $170$~days.
    The fingerprints of the mobile browsers are generally more stable than the fingerprints of the desktop browsers.

\begin{figure}
      \centering
      \includegraphics[width=.7\columnwidth]{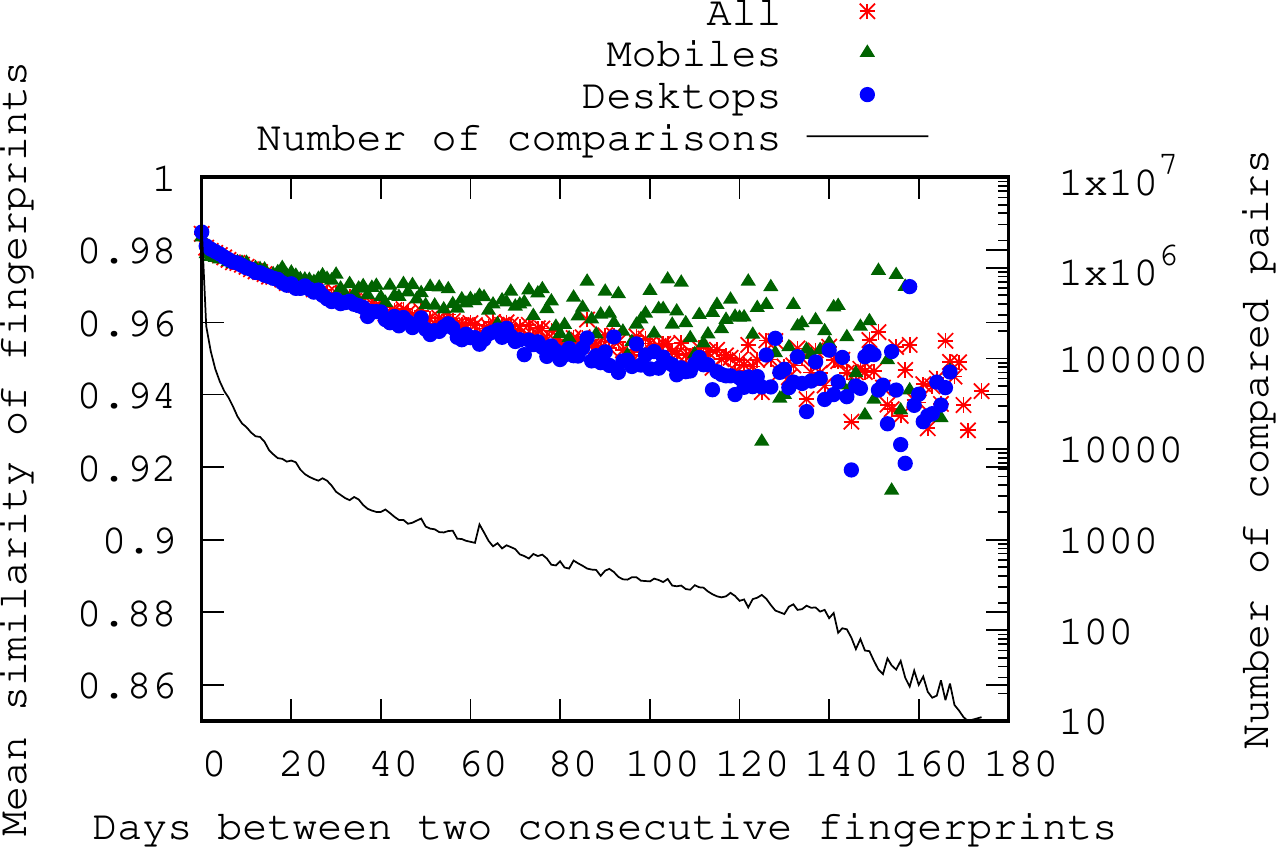}
      \caption{Mean similarity between the consecutive fingerprints in function of the time difference, with the number of compared pairs, for the overall, the mobile, and the desktop groups.}
      \label{fig:stability}
    \end{figure}

    Figure~\ref{fig:stability} displays the mean similarity between the consecutive fingerprints in function of the time difference.
    The ranges~$\Delta$ are expressed in days, so that day~$d$ on the x-axis represents the fingerprints separated by~$\Delta = [d; d+1[$~days.
    We ignore the comparisons of the time ranges that have less than $10$~pairs, or that have a time difference higher than the limit of our experiment ($182$~days).
    These outliers account for less than $0.03$\% of each group.
    The results are obtained by comparing $3,725,373$~pairs for the overall group, $2,912,860$~pairs for the desktop group, and $594,591$~pairs for the mobile group.

  \subsection{Performance}

    \subsubsection{Time Resource Consumption}
      Our probe takes several seconds to collect the attributes that compose our fingerprints.
      The median collection time of our fingerprints is of $2.92$ seconds.
      Mobile browsers take more time to provide the collected attributes, with a median collection time of $4.44$ seconds, against $2.64$ seconds for the desktop browsers.
      This is less than the median loading time of web pages\footnote{
        \url{https://httparchive.org/reports/loading-speed\#ol}
      }, which is of $6.6$~seconds for the desktop browsers, and of $19.6$~seconds for the mobile browsers, at the date of March~$1$, $2020$.

      \begin{figure}
        \centering
        \includegraphics[width=.7\columnwidth]{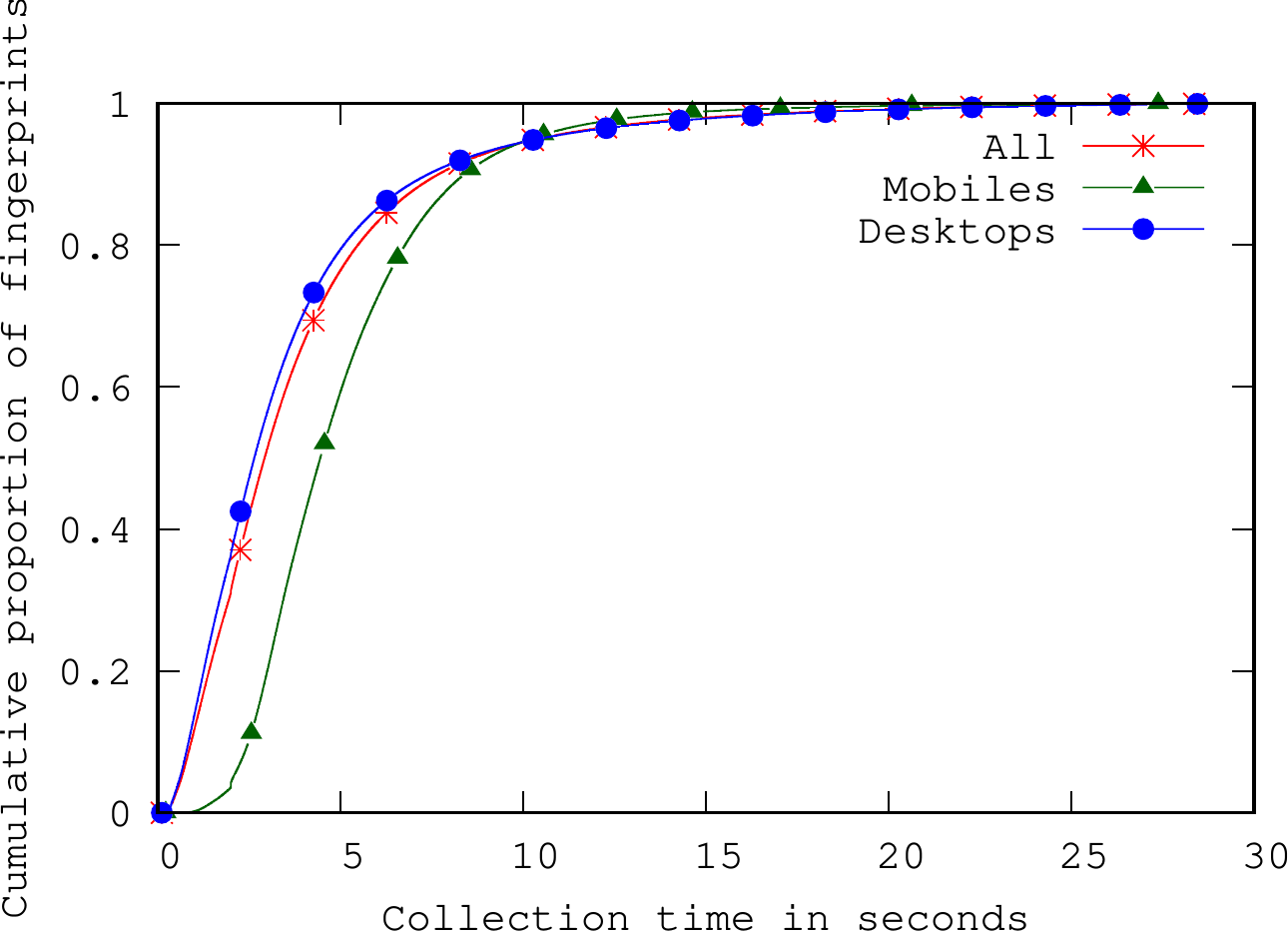}
        \caption{Cumulative distribution of the collection time of the fingerprints in seconds, for the overall, the mobile, and the desktop groups.}
        \label{fig:fps-fetch-time}
      \end{figure}

      Figure~\ref{fig:fps-fetch-time} displays the cumulative distribution of the collection time of our fingerprints.
      We measure the collection time by the time difference between the starting of the script and the fingerprint sending.
      Some values take from several hours to days.
      They can come from a web page that is put in background or accessed after a long time.
      We limit our population to the fingerprints that take less than $30$~seconds to collect, and consider the higher values as outliers.
      The outliers account for less than $1$\%~of each group.

    \subsubsection{Memory Resource Consumption}
      Our script consumes a dozen of kilobytes per fingerprint, a size that is easily handled by the current storage and bandwidth capacities.
      Half of our fingerprints weigh less than $7,550$~bytes, and $99$\%~less than $14$~kilobytes.
      The fingerprints of the desktop browsers are heavier, with $95$\% of the fingerprints weighing less than $12,082$~bytes, against $8,020$~bytes for the fingerprints of mobile browsers.
      This is due to heavy attributes being lighter on mobile browsers, like the list of plugins that is usually empty.

      \begin{figure}
        \centering
        \includegraphics[width=.7\columnwidth]{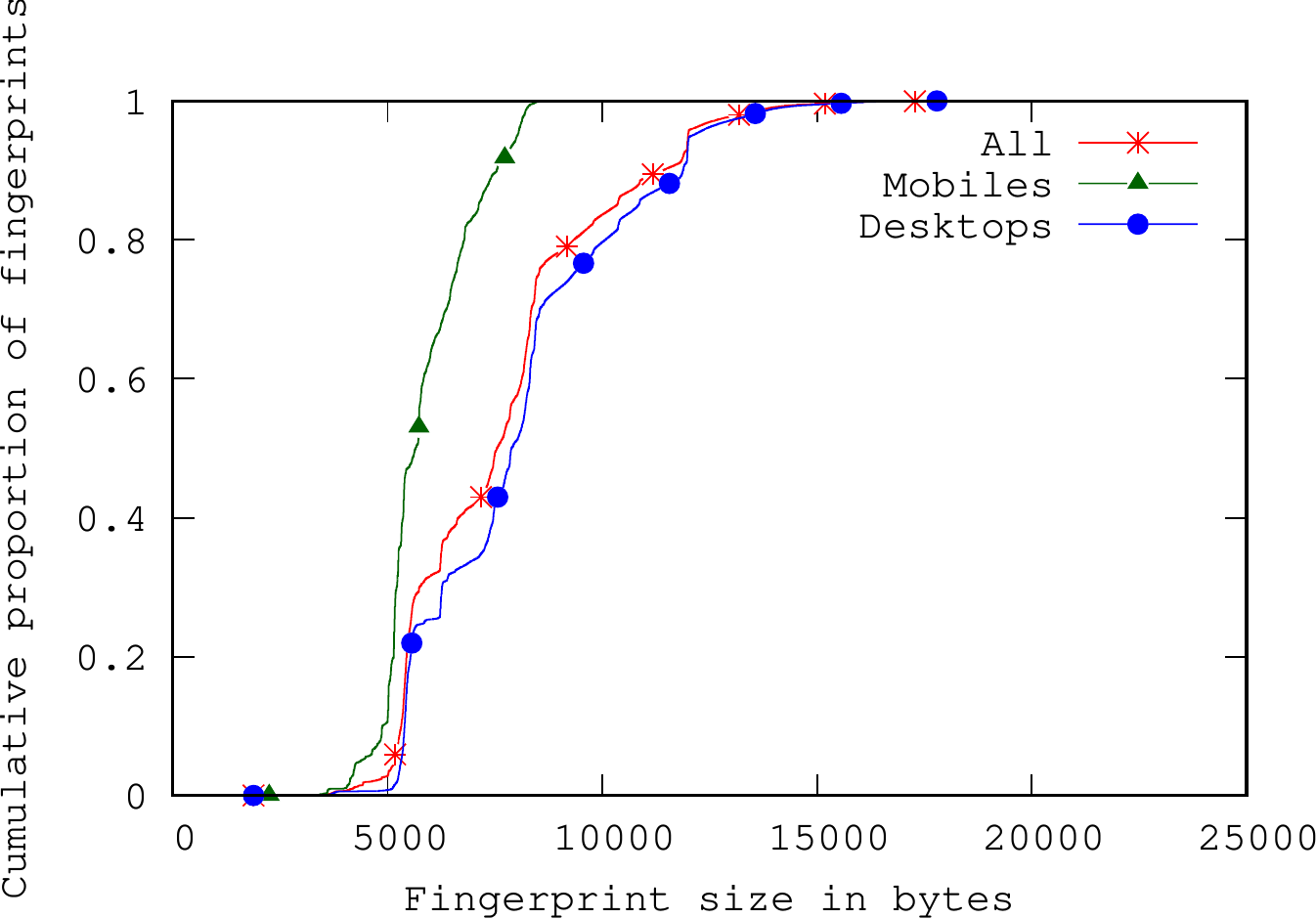}
        \caption{Cumulative distribution of the fingerprint size in bytes, for the overall, the mobile, and the desktop groups.}
        \label{fig:fps-size-cdf}
      \end{figure}

      Figure~\ref{fig:fps-size-cdf} displays the cumulative distribution of the size of our fingerprints in bytes.
      The average fingerprint size is $\mu=7,692$~bytes, and the standard deviation is $\sigma=2,294$.
      We remove $1$~fingerprint from a desktop browser that is considered an outlier because of its size being greater than~$\mu + 15 \cdot \sigma$.

\section{Synthesis of Results and Conclusion}
\label{sec:conclusion}
  In this study, we evaluate the properties offered by browser fingerprints as an additional web authentication factor, through the analysis of a large-scale real-life fingerprint dataset.
  We show that browser fingerprints offer a satisfying \emph{distinctiveness}, as $81$\% of our fingerprints are only shared by one browser.
  Moreover, fingerprints are \emph{stable}.
  At least $90$\%~of our attributes are expected to stay identical between two observations of the fingerprint of a browser, even if the observations are separated by nearly $6$~months.
  We validate that fingerprints offer a high \emph{performance}, as they only weigh a dozen of kilobytes, and take a few seconds to collect.
  We conclude that browser fingerprints provide satisfying properties for an additional web authentication factor, and can strengthen password-based systems without a major loss of usability.

\begin{acknowledgement}
  We want to thank Benoît Baudry and David Gross-Amblard for their valuable comments, together with Alexandre Garel for his work on the experiment.
  This is a postprint of a contribution published in \textit{Innovative Mobile and Internet Services in Ubiquitous Computing}, edited by Leonard Barolli, Aneta Poniszewska-Maranda, and Hyunhee Park, and published by Springer International Publishing.
  The final authenticated version is available online at: \url{https://doi.org/10.1007/978-3-030-50399-4\_16}.
\end{acknowledgement}

\bibliographystyle{spmpsci}
\bibliography{biblio}

\end{document}